\documentclass[debug,overfull,figures]{epl}

\title{Theory
of the propagation of coupled waves
in arbitrarily-inhomogeneous stratified media}
\shorttitle{Propagation of coupled waves
in stratified media}
\author{Kihong Kim\inst{1}\thanks{E-mail:
\email{khkim@ajou.ac.kr}} \and Dong-Hun Lee\inst{2} \and H. Lim\inst{3}}
\shortauthor{K. Kim \etal}
\institute{
  \inst{1} Department of
Molecular Science and Technology, Ajou University, Suwon,
Korea\\
  \inst{2} Department of Astronomy and Space Science, Kyung Hee
University, Yongin, Korea\\
\inst{3} Department of
Electrical Engineering, Ajou University, Suwon, Korea}

\begin{document}

\maketitle

\begin{abstract}
We generalize the invariant imbedding theory of the wave propagation
and derive new invariant imbedding equations for the propagation of
arbitrary number of coupled waves of any kind in
arbitrarily-inhomogeneous stratified media, where the wave equations
are effectively one-dimensional. By doing this, we transform the
original boundary value problem of coupled second-order differential
equations to an initial value problem of coupled first-order
differential equations, which makes the numerical solution of the
coupled wave equations much easier. Using the invariant imbedding
equations, we are able to calculate the matrix reflection and
transmission coefficients and the wave amplitudes inside the
inhomogeneous media exactly and efficiently. We establish the
validity and the usefulness of our results by applying them to the
propagation of circularly-polarized electromagnetic waves in
one-dimensional photonic crystals made of isotropic chiral media. We
find that there are three kinds of bandgaps in these structures and
clarify the nature of these bandgaps by exact calculations.
\end{abstract}

\section{Introduction}
The phenomena of the coupling of two or more wave modes in
inhomogeneous media and mode conversion between them are ubiquitous
in various branches of science, including plasma physics, optics,
condensed matter physics and electrical engineering
\cite{1,budden,yariv,doro,kpw,2}. In this Letter, we develop a
generalization of the powerful invariant imbedding method
\cite{2,kly,bell,ram,hein,kim1,kim2,kim3} to the case of several
coupled waves in stratified media. Starting from a very general wave
equation of a matrix form, we derive a new version of the invariant
imbedding equations for calculating the reflection and transmission
coefficients and the field amplitudes. By doing this, we transform
the original boundary value problem of coupled second-order
differential equations to an initial value problem of coupled
first-order differential equations. This makes the numerical
solution of the coupled wave equations much easier. Furthermore, our
equations have a great advantage that there is no singular
coefficient even in the cases where the material parameters change
discontinuously at the boundaries and inside the inhomogeneous
medium. We check the validity and the usefulness of our invariant
imbedding equations by applying them to the propagation of
electromagnetic waves in stratified chiral media. By calculating the
matrix reflection and transmission coefficients exactly, we clarify
the nature of the three different photonic bandgaps that can exist
in photonic crystals made of chiral media.

\section{Theory}
We consider a system of $N$ coupled waves propagating in a stratified medium,
where all parameters may
depend on only one spatial coordinate. We take this
coordinate as the $z$ axis and assume the inhomogeneous medium
of thickness $L$ lies in $0\le z\le L$.
We also assume that all $N$ waves propagate in the $xz$ plane.
The $x$ component of the wave vector, $q$,
is a constant
and the dependence on $x$ of all wave functions
can be taken as being through a factor $e^{iqx}$.
In a large class of interesting problems,
the wave equation of $N$ coupled waves in the present situation has the form
\begin{eqnarray}
{{d^2 \psi}\over{dz^2}}-\frac{d\cal E}{dz}{\cal E}^{-1}(z)
\frac{d\psi}{dz}+\left[{\cal E}(z)K^2{\cal M}(z)-q^2I\right]\psi=0,
\label{eq:wave}
\end{eqnarray}
where $\psi=(\psi_1,\cdots,\psi_N)^T$ is an $N$-component vector
wave function and ${\cal E}$ and ${\cal M}$ are $N\times N$ matrix
functions that depend on $z$ in an arbitrary manner inside the
inhomogeneous medium. We assume that the waves are incident from the
vacuum region where $z>L$ and transmitted to another vacuum region
where $z<0$. $I$ is a unit matrix and $K$ is a diagonal matrix such
that $K_{ij}=k_i\delta_{ij}$, where $k_i$ is the magnitude of the
vacuum wave vector for the $i$-th wave. $\cal E$ and $\cal M$ are
unit matrices in the vacuum region. The nonsingular functions ${\cal
E}(z)$ and ${\cal M}(z)$, which specify the material properties of
the medium and/or the external conditions, can change
discontinuously at the boundaries and at discrete $z$ values inside
the medium. By assigning ${\cal E}(z)$ and ${\cal M}(z)$ suitably,
eq.~(\ref{eq:wave}) is able to describe many different kinds of
waves in a large number of stratified media.

There are numerous examples where the effective wave equations have
precisely the same form as eq.~(\ref{eq:wave}). Later in this
Letter, we will apply our theory to the propagation of
electromagnetic waves of two different polarizations in layered
chiral media, where $\psi$ is a two-component vector and $\cal E$,
$\cal M$ and $K$ are $2\times 2$ matrices. Another interesting
example is the propagation of the probe and phase-conjugate waves in
layered phase-conjugating media \cite{henk,bla}. A wide variety of
mode conversion phenomena observed in space and laboratory plasmas
can also be studied using eq.~(\ref{eq:wave})
\cite{1,budden,kim3,hinkel}.

Following Gryanik and Klyatskin \cite{2}, we generalize
eq.~(\ref{eq:wave}) slightly, by replacing the vector wave function
$\psi$ by an $N\times N$ matrix wave function $\Psi$, the $j$-th
column vector $(\Psi_{1j},\cdots,\Psi_{Nj})^T$ of which represents
the wave function when the incident wave consists only of the $j$-th
wave. We are interested in the $N\times N$ reflection and
transmission coefficient matrices $r=r(L)$ and $t=t(L)$. Let us
introduce a matrix
\begin{eqnarray}
g(z,z^\prime)=\left\{ \begin{array}{ll}
{\mathcal T}\exp\left[i
\int_{z^\prime}^zdz^{\prime\prime}
~{\cal E}(z^{\prime\prime})P\right],
&~z>z^\prime\\
\tilde{\mathcal T}\exp\left[-i
\int_{z^\prime}^zdz^{\prime\prime}
~{\cal E}(z^{\prime\prime})P\right],
&~z<z^\prime \end{array} \right.
\label{eq:gf}
\end{eqnarray}
where $\mathcal T$ and $\tilde{\mathcal T}$
are the time-ordering and anti-time-ordering operators respectively.
$P$ is a diagonal matrix satisfying $P_{ij}=p_i\delta_{ij}$
and $p_i$ is the negative $z$ component of the vacuum wave vector for the $i$-th
wave.
It is straightforward to prove that $g(z,z^\prime)$ satisfies
the differential equations
\begin{eqnarray}
\frac{\partial}{\partial z}g(z,z^\prime)=i~ {\rm sgn}(z-z^\prime)
~{\mathcal E}(z)Pg(z,z^\prime),
~~\frac{\partial}{\partial z^\prime}g(z,z^\prime)=
-i ~{\rm sgn}(z-z^\prime)~ g(z,z^\prime){\mathcal E}(z^\prime)P.
\label{eq:gf2}
\end{eqnarray}
Using eqs.~(\ref{eq:gf}) and (\ref{eq:gf2}), the wave equation
(\ref{eq:wave}) is transformed to an integral equation
\begin{eqnarray}
&&\Psi(z,L)=g(z,L)\nonumber\\
&&~~-{i\over 2}\int_0^L dz^\prime g(z,z^\prime)\left[
{\cal E}(z^\prime)P-P{\cal M}(z^\prime)-q^2P^{-1}{\cal M}(z^\prime)
+q^2P^{-1}{\cal E}^{-1}(z^\prime)\right]\Psi(z^\prime,L),
\label{eq:inteq}
\end{eqnarray}
where we consider $\Psi$ as a function
of both $z$ and $L$.
We take a partial derivative of this equation with respect to $L$ and obtain
\begin{eqnarray}
&&\frac{\partial \Psi(z,L)}{\partial L}=i\Psi(z,L)\alpha(L)+\Phi(z,L),
\end{eqnarray}
where
\begin{eqnarray}
\alpha(L)={\cal E}(L)P-{1\over 2}\left[{\cal E}(L)P-P{\cal M}(L)
-q^2P^{-1}{\cal M}(L)
+q^2P^{-1}{\cal E}^{-1}(L)\right]\Psi(L,L),
\end{eqnarray}
and $\Phi(z,L)$ satisfies an equation similar to eq.~(\ref{eq:inteq})
except that there is no source term (that is, $g(z,L)$).
This implies $\Phi(z,L)=0$ and then we
have
\begin{equation}
\frac{\partial \Psi(z,L)}{\partial L}=i\Psi(z,L)\alpha(L).
\label{eq:main}
\end{equation}
Taking now the derivative of $\Psi(L,L)$ with respect to $L$,
we obtain
\begin{eqnarray}
\frac{d \Psi(L,L)}{d L}=\frac{\partial \Psi(z,L)}{\partial z}
\Bigg\vert_{z=L}+
\frac{\partial \Psi(z,L)}{\partial L}\Bigg\vert_{z=L}
=i{\cal E}(L)P\left[r(L)-I\right]+i\Psi(L,L)\alpha(L).
\end{eqnarray}
Since $\Psi(L,L)=I+r(L)$, we easily find the ($N\times N$ matrix) invariant imbedding
equation satisfied by $r(L)$:
\begin{eqnarray}
{{dr}\over{dL}}&=&i\left[r(L){\cal E}(L)P+{\cal E}(L)Pr(L)\right]
\nonumber\\
&&-{i \over 2}[r(L)+I]\left[
{\cal E}(L)P-P{\cal M}(L)-q^2P^{-1}{\cal M}(L)
+q^2P^{-1}{\cal E}^{-1}(L)\right][r(L)+I].
\label{eq:imbedr}
\end{eqnarray}

Similarly by setting $z=0$ in eq.~(\ref{eq:main}), we find the
invariant imbedding equation for $t(L)$ ($=\Psi(0,L)$):
\begin{eqnarray}
{{dt}\over{dL}}&=&it(L){\cal E}(L)P
\nonumber\\&&-{i \over 2}t(L)\left[
{\cal E}(L)P-P{\cal M}(L)-q^2P^{-1}{\cal M}(L)
+q^2P^{-1}{\cal E}^{-1}(L)\right][r(L)+I].
\label{eq:imbedt}
\end{eqnarray}
These invariant imbedding equations are supplemented with the initial
conditions, $r(0)=0$ and $t(0)=I$.
For given values of
$P$ and $q$ and for arbitrary matrix functions
${\cal E}(L)$ and ${\cal M}(L)$, we
solve the coupled nonlinear ordinary
differential equations (\ref{eq:imbedr})
and (\ref{eq:imbedt})
numerically using the initial conditions, and
obtain the reflection and transmission coefficient matrices $r$
and $t$ as functions of $L$.
The invariant imbedding method can also be used in calculating
the field amplitude $\Psi(z)$ inside the inhomogeneous medium.
Rewriting eq.~(\ref{eq:main}), we get
\begin{eqnarray}
{{\partial\Psi(z,l)}\over{\partial l}}&=&i\Psi(z,l){\cal E}(l)P
\nonumber\\&&
-{i \over 2}\Psi(z,l)\left[
{\cal E}(l)P-P{\cal M}(l)-q^2P^{-1}{\cal M}(l)
+q^2P^{-1}{\cal E}^{-1}(l)\right][r(l)+I].
\label{eq:imbedf}
\end{eqnarray}
For a given $z$ ($0<z<L$), the field amplitude $\Psi(z,L)$
is obtained by integrating this equation from $l=z$ to $l=L$
using the initial condition $\Psi(z,z)=I+r(z)$.

\section{Application}
Eqs.~(\ref{eq:imbedr}), (\ref{eq:imbedt}) and (\ref{eq:imbedf}),
which have never been derived before to the best of our knowledge,
will be the starting point in our future analysis
of a variety of wave coupling and mode conversion phenomena.
In the rest of this Letter, we establish the validity and the utility
of our invariant imbedding equations
by applying them to the problem of the electromagnetic wave
propagation in stratified chiral media.

Isotropic chiral media are those where the appropriate constitutive
relations are given by
\begin{eqnarray}
{\bf D}=\epsilon{\bf E}+i\gamma{\bf H},~~~
{\bf B}=\mu{\bf H}-i\gamma{\bf E}.
\label{eq:sil}
\end{eqnarray}
The parameters $\epsilon$, $\mu$ and $\gamma$ are the dielectric
permittivity, the magnetic permeability and the chiral index
respectively \cite{lindell,sil,3}. Some researchers use alternative
constitutive relations \cite{4,jag}
\begin{eqnarray}
{\bf D}=\tilde\epsilon {\bf E}+i\xi{\bf B},~~~
{\bf H}={\bf B}/\mu+i\xi{\bf E}.
\end{eqnarray}
The two relations give identical results if
the parameters are identified by
\begin{eqnarray}
\tilde\epsilon=\epsilon-\gamma^2/\mu, ~~~\xi=\gamma/\mu.
\label{eq:bpe}
\end{eqnarray}
We will use eq.~(\ref{eq:sil}) from now on. In recent years, there
have been a large number of theoretical
\cite{lindell,shivola,3,4,jag,5,6} and experimental
\cite{sil,luk,sil2,gen1,gen2} studies on the wave propagation in
various kinds of chiral media.

From the Maxwell's equations and the constitutive relations, we are
able to derive the wave equations satisfied by the electric field in
inhomogeneous chiral media:
\begin{eqnarray}
&&\mu\nabla\times\left({1\over\mu}\nabla\times{\bf E}\right)
=\left(\epsilon\mu-\gamma^2\right){\omega^2\over c^2}{\bf E}
+{\omega\over c}\left[\gamma\nabla\times{\bf E}+\mu\nabla\times\left(
{\gamma\over\mu}{\bf E}\right)\right].
\label{eq:cwe}
\end{eqnarray}
In the uniform case, right- and left-circularly-polarized waves
are eigenmodes of this equation with the effective refractive indices
$\sqrt{\epsilon\mu}+\gamma$ and $\sqrt{\epsilon\mu}-\gamma$ respectively.
In inhomogeneous media, these two modes are no longer eigenmodes and
are coupled to each other.
The equation satisfied by the magnetic field $\bf H$
is similar except that the roles of $\epsilon$ and $\mu$ are reversed.
In media stratified in the $z$ direction, $\epsilon$,
$\mu$ and $\gamma$ are functions of $z$ only.
For plane waves propagating in the $xz$-plane, the $x$ dependence
of all field components is contained in the factor $e^{iqx}$.
In this situation, we can eliminate $E_x$, $E_z$, $H_x$ and $H_z$
from eq.~(\ref{eq:cwe}) and obtain two coupled wave equations
satisfied by $E_y=E_y(z)$ and $H_y=H_y(z)$,
which turn out to have precisely
the same form as eq.~(\ref{eq:wave}) with
\begin{eqnarray}
\psi=\pmatrix{ E_y \cr H_y \cr}, ~~
K=\pmatrix{k&0\cr 0&k\cr},~~{\cal E}=\pmatrix{\mu&i\gamma\cr -i\gamma&\epsilon\cr},~~
{\cal M}=\pmatrix{\epsilon&i\gamma\cr -i\gamma&\mu\cr},
\label{eq:matrix}
\end{eqnarray}
where $k=\omega/c$.

We have used eqs.~(\ref{eq:imbedr}), (\ref{eq:imbedt}) and (\ref{eq:matrix})
in calculating the reflection and transmission coefficients in various situations.
In all cases where exact solutions by other methods are available,
our theory gives the same results. In our notation,
$r_{11}$($r_{21}$) is the reflection coefficient
when the incident wave is $s$-polarized and the reflected wave is $s$($p$)-polarized.
Similarly, $r_{22}$($r_{12}$) is the reflection coefficient
when the incident wave is $p$-polarized and the reflected wave is $p$($s$)-polarized.
Similar definitions are applied to the transmission coefficients.
By a suitable linear combination of these coefficients, we are able to obtain
a new set of the reflection and transmission coefficients $r_{ij}$ and $t_{ij}$, where
$i$ and $j$ are either $+$ or $-$ \cite{3}.
For instance, $r_{++}$($r_{-+}$) represents the reflection
coefficient when the incident wave is right-circularly-polarized
and the reflected wave is right(left)-circularly-polarized.
The reflectances and transmittances
are defined by $R_{ij}=\vert r_{ij}\vert^2$ and $T_{ij}=\vert t_{ij}\vert^2$.

As an example, we consider a uniform chiral layer of finite
thickness with the parameters $\epsilon$, $\mu$ and $\gamma$, placed
between uniform achiral media of infinite thicknesses. In this case,
the electromagnetic wave equations can be solved analytically,
following the methods used in elementary quantum mechanics. Lekner
has presented an exact analytical solution of this problem
\cite{3}\footnote{There are three typos in Lekner's solution. In the
expressions of $G_1^{\pm}$ and $G_2^{\pm}$ in eq.~(A2),
$c_1^2+c_+c_-$ and $c_2^2+c_+c_-$ have to be replaced by
$c_1^2-c_+c_-$ and $c_2^2-c_+c_-$ respectively. The expression
$Z_+Z_-^2$ appearing at the end of the equation for $t_{\rm ps}$ in
eq.~(A5) has to be replaced by $Z_+^2Z_-$.}. In defining the
reflection and transmission coefficients, $r_{\rm ss}$, $r_{\rm
sp}$, $r_{\rm ps}$, $r_{\rm pp}$, $t_{\rm ss}$, $t_{\rm sp}$,
$t_{\rm ps}$ and $t_{\rm pp}$, Lekner uses different conventions
from ours. In order to compare his solution with ours, we need to
identify $r_{11}=r_{\rm ss}$, $r_{12}=r_{\rm ps}$, $r_{21}=-r_{\rm
sp}$, $r_{22}=-r_{\rm pp}$, $t_{11}=t_{\rm ss}$, $t_{12}=t_{\rm
ps}$, $t_{21}=t_{\rm sp}$ and $t_{22}=t_{\rm pp}$, if the magnetic
permeability of the media outside the chiral layer is equal to 1.
Taking these into account, we have verified analytically that
Lekner's expressions for the reflection and transmission
coefficients satisfy our invariant imbedding equations exactly.

\begin{figure}
\onefigure[width=8cm]{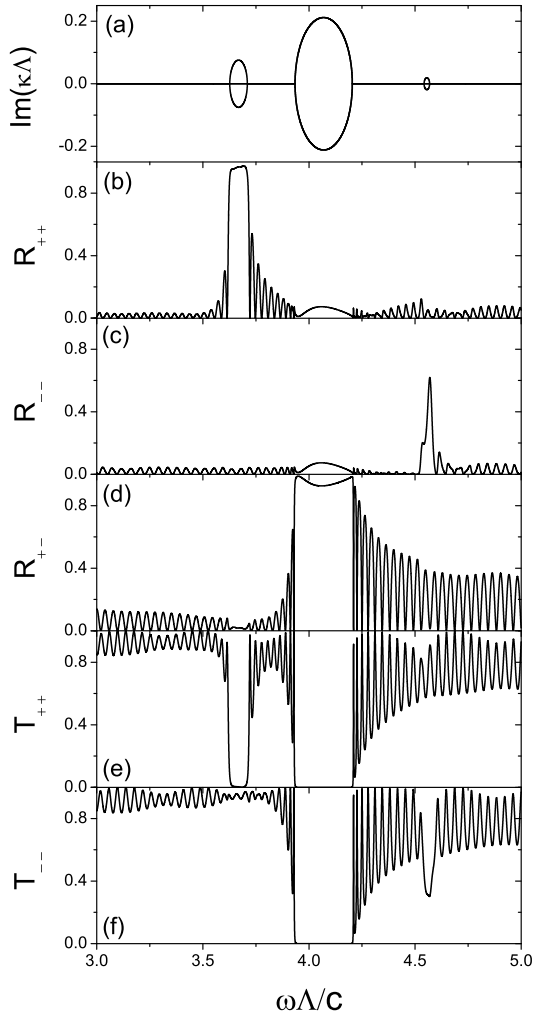} \caption{(a) Imaginary part of the
Bloch wave number $\kappa$ for an infinitely large photonic crystal
made of alternating layers of chiral and dielectric materials of the
same thicknesses $\Lambda/2$. The chiral layer has the parameter
values of $\epsilon=4$, $\mu=1$ and $\gamma=0.3$ and the dielectric
layer has $\epsilon=2$, $\mu=1$ and $\gamma=0$. The transverse
component of the wave vector, $q$, is given by
$q=\omega\sin\theta/c$, where $\theta=45^\circ$. The $z$ component
of the vacuum wave vector matrix, $P$, is given by $P=pI$, where
$p=\omega\cos\theta/c$. The frequency region where ${\rm Im}~
\kappa$ is nonzero corresponds to a bandgap.
 (b-f) Reflectance and transmittance spectra
for a one-dimensional photonic crystal made of alternating layers of
chiral and dielectric materials. The parameter values and the values
of $\theta$, $p$ and $q$ are the same as in (a) and the number of
periods is 50.} \label{f.1}
\end{figure}

In fig.~\ref{f.1}, we plot the reflectances $R_{++}$, $R_{--}$ and
$R_{+-}$ and the transmittances $T_{++}$ and $T_{--}$, when a wave
is incident at $\theta=45^\circ$ on a one-dimensional photonic
crystal made of alternating chiral and dielectric layers of the same
thicknesses $\Lambda/2$. It can be proved easily that
$R_{12}=R_{21}$ and $R_{+-}=R_{-+}$. The chiral layer has the
parameter values of $\epsilon=4$, $\mu=1$ and $\gamma=0.3$ and the
dielectric layer has $\epsilon=2$, $\mu=1$ and $\gamma=0$. The total
number of periods is 50. The $x$ component of the wave vector, $q$,
is given by $q=\omega\sin\theta/c$ and the $z$ component of the
vacuum wave vector matrix, $P$, is given by $P=pI$, where
$p=\omega\cos\theta/c$. Also plotted is the imaginary part of the
Bloch wave number $\kappa$ for an infinitely large photonic crystal.
This quantity was obtained using an exact analytical expression for
the dispersion relation of infinitely large photonic crystals made
of two different kinds of alternating chiral layers, which we have
derived recently \cite{yoo}. The frequency region where the
imaginary part of $\kappa$ is nonzero corresponds to a photonic
bandgap.

We find an excellent agreement between the analytical result on
the dispersion relation and the reflectance and transmittance spectra.
In general, there are three kinds of bandgaps, two of which are
so-called co-polarization bandgaps and one of which is called a
cross-polarization bandgap. Unlike in previous studies of this
phenomenon \cite{5,6}, our theory is free of any approximation
and provides exact band structures.
For large values of $\gamma$ and $\theta$,
these three bandgaps can be well-separated, as demonstrated in fig.~\ref{f.1},
where we show the second group of bandgaps.
The $R_{+-}$ spectrum clearly displays a cross-polarization bandgap
and the $R_{++}$ and $R_{--}$ spectra show co-polarization bandgaps.
The transmittance spectra show that a right(left)-circularly-polarized
wave, the frequency of which lies in the co-polarization bandgap
of left(right)-circularly-polarized waves, is freely transmitted.

It is straightforward to apply our method to more general situations where
the parameters $\epsilon$, $\mu$ and $\gamma$ are arbitrary functions of $z$.
For example, we can easily study the effects of defects and
randomness on the wave propagation
in chiral media using
eqs.~(\ref{eq:imbedr}), (\ref{eq:imbedt}), (\ref{eq:imbedf}) and (\ref{eq:matrix}).
Our equations can also be applied to the cases where both $\epsilon$
and $\mu$ take negative values with no modification.
A detailed study of this so-called {\it negative refractive
index} medium \cite{pendry}, which is also chiral,
is of great interest and will be presented elsewhere.
We have also applied our method successfully to
a number of other coupled wave problems,
such as the phase-conjugate reflection of light from nonlinear phase-conjugating media,
the light propagation in uniaxial and biaxial media and the mode conversion phenomena
in both unmagnetized and magnetized plasmas \cite{future}.
In these studies, we solve the full wave equations exactly, without using common
approximations such as the slowly varying envolope approximation and the WKB
approximation.
These results will be presented in a near future.

\acknowledgments This work has been supported by the KOSEF through
grant number R14-2002-062-01000-0. D.-H. Lee was supported partially
by Kyung Hee University.

\end{document}